\documentclass[twocolumn,eqsecnum,aps,floatfix]{revtex4} 

\usepackage{graphicx}
\usepackage{dcolumn}
\usepackage{amsmath}
\usepackage{amssymb}
\usepackage{epstopdf}

\makeatletter
\def\btt#1{\texttt{\@backslashchar#1}}%
\DeclareRobustCommand\bblash{\btt{\@backslashchar}}%
\makeatother

\begin{document}
\title{Solar neutrino measurements in Super--Kamiokande--II}
\newcounter{foots}
\newcounter{notes}
\newcommand{\authoraticrr}{$^{1}$}
\newcommand{\authoraticrracckek}{$^{1,\textrm{a}}$}
\newcommand{\authoratncen}{$^{2}$}
\newcommand{\authoratipmuncen}{$^{2,3}$}
\newcommand{\authoratipmu}{$^{3}$}
\newcommand{\authoratipmuicrr}{$^{1,3}$}
\newcommand{\authoratipmuuci}{$^{6,3}$}
\newcommand{\authoratbu}{$^{4}$}
\newcommand{\authoratbupenn}{$^{4,\textrm{b}}$}
\newcommand{\authoratbnl}{$^{5}$}
\newcommand{\authoratuci}{$^{6}$}
\newcommand{\authoratcsu}{$^{7}$}
\newcommand{\authoratcnu}{$^{8}$}
\newcommand{\authoratduke}{$^{9}$}
\newcommand{\authoratgmu}{$^{10}$}
\newcommand{\authoratgifu}{$^{11}$}
\newcommand{\authoratuh}{$^{12}$}
\newcommand{\authoratui}{$^{13}$}
\newcommand{\authoratkanagawa}{$^{14}$}
\newcommand{\authoratkek}{$^{15}$}
\newcommand{\authoratkekicrr}{$^{15,1}$}
\newcommand{\authoratkekkashiwa}{$^{15}$}
\newcommand{\authoratkobe}{$^{16}$}
\newcommand{\authoratkyoto}{$^{17}$}
\newcommand{\authoratkyototriumf}{$^{17,\textrm{c}}$}
\newcommand{\authoratlanluci}{$^{18,6}$}
\newcommand{\authoratlsu}{$^{19}$}
\newcommand{\authoratumd}{$^{20}$}
\newcommand{\authoratduluth}{$^{21}$}
\newcommand{\authoratmiyagi}{$^{22}$}
\newcommand{\authoratnagoya}{$^{23}$}
\newcommand{\authoratsuny}{$^{24}$}
\newcommand{\authoratniigata}{$^{25}$}
\newcommand{\authoratokayama}{$^{26}$}
\newcommand{\authoratosaka}{$^{27}$}
\newcommand{\authoratseoul}{$^{28}$}
\newcommand{\authoratshizuoka}{$^{29}$}
\newcommand{\authoratshizuokaseika}{$^{30}$}
\newcommand{\authoratskku}{$^{31}$}
\newcommand{\authorattohoku}{$^{32}$}
\newcommand{\authorattokai}{$^{33}$}
\newcommand{\authorattit}{$^{34}$}
\newcommand{\authorattokyo}{$^{35}$}
\newcommand{\authorattsinghua}{$^{36}$}
\newcommand{\authoratwarsaw}{$^{37}$}
\newcommand{\authoratwarsawuci}{$^{37,6}$}
\newcommand{\authoratuw}{$^{38}$}
\newcommand{\authoratuwduluth}{$^{38,20}$}
\newcommand{\addressoficrr}[1]{$^{1}$ #1 }
\newcommand{\addressofncen}[1]{$^{2}$ #1 }
\newcommand{\addressofipmu}[1]{$^{3}$ #1 }
\newcommand{\addressofbu}[1]{$^{4}$ #1 }
\newcommand{\addressofbnl}[1]{$^{5}$ #1 }
\newcommand{\addressofuci}[1]{$^{6}$ #1 }
\newcommand{\addressofcsu}[1]{$^{7}$ #1 }
\newcommand{\addressofcnu}[1]{$^{8}$ #1 }
\newcommand{\addressofduke}[1]{$^{9}$ #1 }
\newcommand{\addressofgmu}[1]{$^{10}$ #1 }
\newcommand{\addressofgifu}[1]{$^{11}$ #1 }
\newcommand{\addressofuh}[1]{$^{12}$ #1 }
\newcommand{\addressofui}[1]{$^{13}$ #1 }
\newcommand{\addressofkanagawa}[1]{$^{14}$ #1 }
\newcommand{\addressofkek}[1]{$^{15}$ #1 }
\newcommand{\addressofkobe}[1]{$^{16}$ #1 }
\newcommand{\addressofkyoto}[1]{$^{17}$ #1 }
\newcommand{\addressoflanl}[1]{$^{18}$ #1 }
\newcommand{\addressoflsu}[1]{$^{19}$ #1 }
\newcommand{\addressofumd}[1]{$^{20}$ #1 }
\newcommand{\addressofduluth}[1]{$^{21}$ #1 }
\newcommand{\addressofmiyagi}[1]{$^{22}$ #1 }
\newcommand{\addressofnagoya}[1]{$^{23}$ #1 }
\newcommand{\addressofsuny}[1]{$^{24}$ #1 }
\newcommand{\addressofniigata}[1]{$^{25}$ #1 }
\newcommand{\addressofokayama}[1]{$^{26}$ #1 }
\newcommand{\addressofosaka}[1]{$^{27}$ #1 }
\newcommand{\addressofseoul}[1]{$^{28}$ #1 }
\newcommand{\addressofshizuoka}[1]{$^{29}$ #1 }
\newcommand{\addressofshizuokaseika}[1]{$^{30}$ #1 }
\newcommand{\addressofskku}[1]{$^{31}$ #1 }
\newcommand{\addressoftohoku}[1]{$^{32}$ #1 }
\newcommand{\addressoftokai}[1]{$^{33}$ #1 }
\newcommand{\addressoftit}[1]{$^{34}$ #1 }
\newcommand{\addressoftokyo}[1]{$^{35}$ #1 }
\newcommand{\addressoftsinghua}[1]{$^{36}$ #1 }
\newcommand{\addressofwarsaw}[1]{$^{37}$ #1 }
\newcommand{\addressofuw}[1]{$^{38}$ #1 }
\def\pennnow{$\textrm{a}$}
\def\kashiwanow{$\ddagger$}
\def\triumfnow{\S}
\author{
\vspace{0.1cm}
%
J.P.~Cravens\authoratuci,
%
K.~Abe\authoraticrr,
T.~Iida\authoraticrr,
K.~Ishihara\authoraticrr,
J.~Kameda\authoraticrr,
Y.~Koshio\authoraticrr,
A.~Minamino\authoraticrr,
C.~Mitsuda\authoraticrracckek,
M.~Miura\authoraticrr,
S.~Moriyama\authoraticrr
% ...
\\
\smallskip
(The Super-Kamiokande Collaboration) \\ 
\smallskip
\footnotesize
\it
\addressoficrr{Kamioka Observatory, Institute for Cosmic Ray Research, 
University of Tokyo, Kamioka, Gifu, 506-1205, Japan}\\
\addressofncen{Research Center for Cosmic Neutrinos, Institute for Cosmic 
Ray Research, University of Tokyo, Kashiwa, Chiba 277-8582, Japan}\\
\addressofipmu{Institute for the Physics and Mathematics of the Universe, 
University of Tokyo, Kashiwa, Chiba 277-8582, Japan}\\
\addressofbu{Department of Physics, Boston University, Boston, MA 02215, 
USA}\\
\addressofbnl{Physics Department, Brookhaven National Laboratory, Upton, 
NY 11973, USA}\\
\addressofuci{Department of Physics and Astronomy, University of 
California, Irvine, Irvine, CA 92697-4575, USA }\\
\addressofcsu{Department of Physics, California State University, 
Dominguez Hills, Carson, CA 90747, USA}\\
\addressofcnu{Department of Physics, Chonnam National University, Kwangju 
500-757, Korea}\\
\addressofduke{Department of Physics, Duke University, Durham, NC 27708, 
USA} \\
\addressofgmu{Department of Physics, George Mason University, Fairfax, VA 
22030, USA }\\
\addressofgifu{Department of Physics, Gifu University, Gifu, Gifu 
501-1193, Japan}\\
\addressofuh{Department of Physics and Astronomy, University of Hawaii, 
Honolulu, HI 96822, USA}\\
\addressofui{Department of Physics, Indiana University, Bloomington,
  IN 47405-7105, USA} \\
\addressofkanagawa{Physics Division, Department of Engineering, Kanagawa University, Kanagawa, Yokohama 221-8686, Japan} \\
\addressofkek{High Energy Accelerator Research Organization (KEK), 
Tsukuba, Ibaraki 305-0801, Japan }\\
\addressofkobe{Department of Physics, Kobe University, Kobe, Hyogo 
657-8501, Japan}\\
\addressofkyoto{Department of Physics, Kyoto University, Kyoto 606-8502, 
Japan}\\
\addressoflanl{Physics Division, P-23, Los Alamos National Laboratory, Los 
Alamos, NM 87544, USA }\\
\addressoflsu{Department of Physics and Astronomy, Louisiana State 
University, Baton Rouge, LA 70803, USA }\\
\addressofumd{Department of Physics, University of Maryland, College Park, 
MD 20742, USA }\\
\addressofduluth{Department of Physics, University of Minnesota, Duluth, 
MN 55812-2496, USA}\\
\addressofmiyagi{Department of Physics, Miyagi University of Education, 
Sendai, Miyagi 980-0845, Japan}\\
\addressofnagoya{Solar Terrestrial Environment Laboratory, Nagoya University, Nagoya, Aichi 
464-8602, Japan}\\
\addressofsuny{Department of Physics and Astronomy, State University of 
New York, Stony Brook, NY 11794-3800, USA}\\
\addressofniigata{Department of Physics, Niigata University, Niigata, 
Niigata 950-2181, Japan }\\
\addressofokayama{Department of Physics, Okayama University, Okayama, 
Okayama 700-8530, Japan} \\
\addressofosaka{Department of Physics, Osaka University, Toyonaka, Osaka 
560-0043, Japan}\\
\addressofseoul{Department of Physics, Seoul National University, Seoul 
151-742, Korea}\\
\addressofshizuoka{Department of Systems Engineering, Shizuoka University, 
Hamamatsu, Shizuoka 432-8561, Japan}\\
\addressofshizuokaseika{Department of Informatics in Social Welfare, Shizuoka University 
of Welfare, Yaizu, Shizuoka, 425-8611, Japan}\\
\addressofskku{Department of Physics, Sungkyunkwan University, Suwon 
440-746, Korea}\\
\addressoftohoku{Research Center for Neutrino Science, Tohoku University, 
Sendai, Miyagi 980-8578, Japan}\\
\addressoftokai{Department of Physics, Tokai University, Hiratsuka, 
Kanagawa 259-1292, Japan}\\
\addressoftit{Department of Physics, Tokyo Institute for Technology, 
Meguro, Tokyo 152-8551, Japan }\\
\addressoftokyo{The University of Tokyo, Tokyo 113-0033, Japan }\\
\addressoftsinghua{Department of Engineering Physics, Tsinghua University, Beijing, 100084, China}\\
\addressofwarsaw{Institute of Experimental Physics, Warsaw University, 
00-681 Warsaw, Poland }\\
\addressofuw{Department of Physics, University of Washington, Seattle, WA 
98195-1560, USA}
}
\date{\today}

\begin{abstract}
The results of the second phase of the Super-Kamiokande solar neutrino measurement are presented and compared to the first phase.  The solar neutrino flux spectrum and time-variation as well as oscillation results are statistically consistent with the first phase and do not show spectral distortion.  The time-dependent flux measurement of the combined first and second phases coincides with the full period of solar cycle 23 and shows no correlation with solar activity.  The measured $^8$B total flux is $(2.38 \pm 0.05 (\textrm{stat.}) ^{+0.16}_{-0.15} (\textrm{sys.})) \times 10^ 6~\textrm{cm}^{-2} \textrm{sec}^{-1}$ and the day-night difference is found to be $(-6.3 \pm 4.2 (\textrm{stat.}) \pm 3.7 (\textrm{sys.}))\%$.  There is no evidence of systematic tendencies between the first and second phases.
\end{abstract}

\maketitle

\section{Introduction}
The first phase of the Super-Kamiokande experiment, SK-I [1-2], yielded high precision measurements of the solar neutrino flux.  In spite of the loss of numerous photomultipler tubes [PMT] sustained in an accident, SK continued to collect data with reduced photo-cathode coverage and a higher energy threshold.  Data collection and analysis methods had to be revised due to the loss of detector sensitivity.  Super-Kamiokande's second phase [SK-II] ran from December 2002 to October 2005.

Throughout this paper, the methods and results of SK-II are compared with SK-I and, when differing, are detailed for SK-II.

\section{SK-II Performance}
\subsection{Detector Simulation}
It was determined that the November 12th, 2001 accident sustained by the SK detector was caused by a propagating shock wave initiated by an imploding PMT located at the bottom of the inner detector.  Therefore, blast shields were installed to protect the PMTs against such a chain reaction.  These shields are 1.0 cm-thick transparent acrylic domes allowing light to pass to the PMTs' photo-sensitive surface.  This presents an additional medium through which Cherenkov light must travel.  Reflection and refraction of light on the acrylic surface is accounted for in the SK-II GEANT 3 Monte Carlo detector simulation.  The acrylic shields' transparency at normal incidence is better than 98\% above 400 nm in wavelength.  It is about 86\% at 300 nm.

\begin{figure}[htbp]
\begin{center}
\includegraphics[scale=0.34]{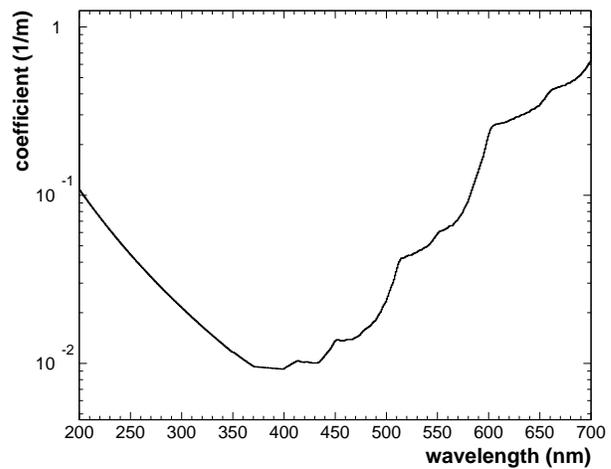}
\caption{Wavelength dependence of the water parameter combined SK-II absorption-reflection coefficient.}
\label{fig:absorption}
\end{center}
\end{figure}

For light propagation in water, both SK-I and SK-II adopt a 3-part model of light attenuation consisting of Rayleigh scattering, Mie scattering, and absorption.  We consider two types of absorption: long wavelength ($\lambda >$ 350 nm) and short wavelength ($\lambda \le$ 350 nm) absorption.  In the long wavelength absorption region, we utilize an independent model derived using direct measurements from an integrating chamber absorption meter [ICAM] applied to pure water~\cite{icam}.  Scattering coefficients and absolute short wavelength absorption are tuned to reproduce energy distributions in LINAC calibration data (see section II.C for a description of the LINAC data).  At short wavelengths, the SK-I model varies the absorption coefficient to describe the changing SK in-tank water transparency as measured by decay electrons from cosmic-ray muons.  In SK-II, the best description has no short wavelength absorption but increased scattering.  Figure~\ref{fig:absorption} shows the attenuation model in both short and long wavelengths for SK-II.

In determining the expected solar neutrino flux spectrum for a range of oscillation parameters, SK-II follows the method of SK-I: the total $^8$B and hep flux values of the BP2004 Standard Solar Model (SSM)~\cite{ssm} are used with the neutrino spectrum based on the $\beta$-delayed $\alpha$ spectrum of $^8$B decay by Ortiz~\cite{ortiz} to calculate the flux of a particular energy bin.  The uncertainties of the neutrino spectrum are taken from Bahcall~\cite{bpspc}.

\subsection{Event Reconstruction}
\subsubsection{Vertex}
The determination of event vertex, direction, and energy with the reduced light collection capability of SK-II has prompted the development of new reconstruction methods.  For vertex reconstruction, the efficiency of the SK-I standard vertex fit significantly drops at energies below the SK-I analysis threshold of 5.0 MeV.  With 40\% photocathode coverage, this corresponds to roughly 25 PMT signal hits.  At 19\% coverage in SK-II, 25 hits translates to 8 MeV.

The timing residual in an event is defined as the time difference between a PMT's hit time $t_i$ and the emission time $t_0$ (fitted to minimize all timing residuals) minus the time it would take Cherenkov light to reach that PMT given the event's vertex $\vec{v}$ in the tank:
\begin{equation} \label{eq:t_residual}
t_{residual} = (t_i - t_0) - |\vec{v} - \vec{h}_i|/c,
\end{equation}
where $\vec{h}_i$ is the vector location of the hit PMT and $c$ is the group velocity of light in water.

In SK-I, vertex reconstruction is accomplished by selecting a limited number of hit PMTs from an event (to reduce bias from PMT dark noise and scattered light hits) and calculating a goodness relation based on the timing residuals of the selected hits and a candidate vertex $\vec{v}$.  A systematic grid search of candidate vertices is performed until the goodness reaches a maximum value.  After that, the vertex position is fine-tuned to further maximize the goodness.  The SK-I reconstruction will not attempt a vertex fit for less than 10 hits.

In contrast, the SK-II reconstruction uses all hits from an event to form the timing residuals for determination of the vertex position.  Bias from PMT dark noise is reduced by constructing a likelihood describing the shape of the timing residual distribution from LINAC calibration data.  This likelihood is then maximized from a vertex search based not on a grid pattern but from a list of vertex candidates calculated from PMT hit combinations of 4 hits each.  The four-hit combinations each define a unique vertex given their timing constraints.  Any event with four hits or more is reconstructed.

\begin{figure}[htbp]
\begin{center}
\includegraphics[scale=0.25]{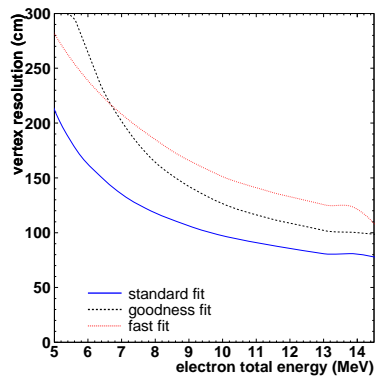}
\caption{Vertex resolution (defined as 68.2\% of reconstructed events which reconstruct inside a sphere of radius $\sigma$ from the correct vertex) of $^8$B Monte Carlo events as a function of total recoil electron energy.}
\label{fig:bonsai_resolution}
\end{center}
\end{figure}

SK-II also makes use of the SK-I goodness-grid search method in its online and initial offline analysis for filtering background events.  The final reconstruction, or the standard fit based on the residual likelihood method, is the final determination of vertex position and can also be seen as a correction for any misreconstructed events which survived the filtering process.  Figure~\ref{fig:bonsai_resolution} shows the various vertex resolutions for the SK-II vertex reconstruction.  The uncertainty of the measured solar neutrino rate due to systematic shifts in vertex position is estimated to be 1.1\%.

Both SK-I and SK-II utilize a fast fit online reconstruction method for pre-filtering low energy events.  Details can be found elsewhere~\cite{sk_full_paper}.

\subsubsection{Direction}
\begin{figure}[htbp]
\begin{center}
\includegraphics[scale=0.35]{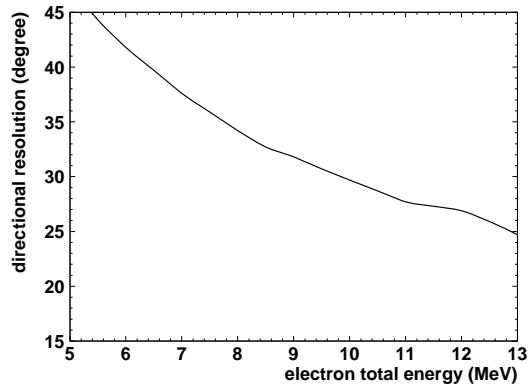}
\caption{Directional resolution of Monte Carlo events as a function of recoil electron total energy.}
\label{fig:direction_resolution}
\end{center}
\end{figure}

The direction reconstruction is identical to the SK-I method: a likelihood function is used to compare Cherenkov ring patterns between data and MC distributions.  Opening angles between the particle direction and reconstructed vertex-to-hit PMT position are scanned using grid searches at varying levels of precision.  The SK-II standard fitter is used to determine the vertex.  The absolute angular resolution (defined as the maximum angular difference between 68\% of the reconstructed and true event directions as determined by MC) of SK-II differs from SK-I by about 10\% and is mostly limited not by detector coverage but by multiple scattering of electrons in the tank.  The difference between data and MC angular resolutions, however, is greater in SK-II due to the larger discrepancies in energy scales between data and MC.  Therefore, we assign an angular resolution systematic error of 6.0\%.  Angular resolution is shown in Figure~\ref{fig:direction_resolution}.

\subsubsection{Energy}
The reconstruction of event energy is identical to that of SK-I with modification specific to SK-II (photo-cathode coverage, blast shields, etc.).  From the number of in-time hit PMTs (coincident within 50 ns) from an event ($N_{\mbox{\tiny{\textit{hit}}}}$), various corrections are made.  The resulting effective hit sum has a consistent value throughout the detector for a given event ($N_{\mbox{\tiny{\textit{eff}}}}$).  From $N_{\mbox{\tiny{\textit{eff}}}}$, we determine energy.  Refer to~\cite{sk_full_paper} for specific information on the conversion from $N_{\mbox{\tiny{\textit{hit}}}}$ to $N_{\mbox{\tiny{\textit{eff}}}}$.

The $N_{\mbox{\tiny{\textit{eff}}}}$-to-energy conversion function must be modified for SK-II due to smaller values of $N_{\mbox{\tiny{\textit{eff}}}}$ corresponding to equivalent energies with larger $N_{\mbox{\tiny{\textit{eff}}}}$ in SK-I.  This is done by generating MC events at discrete input energies between 5 and 80 MeV, calculating their $N_{\mbox{\tiny{\textit{eff}}}}$ values, and then interpolating the energy function.  As with SK-I, energy refers to total energy of the event (kinetic energy plus electron rest mass).

\begin{figure}[htbp]
\begin{center}
\includegraphics[scale=0.35]{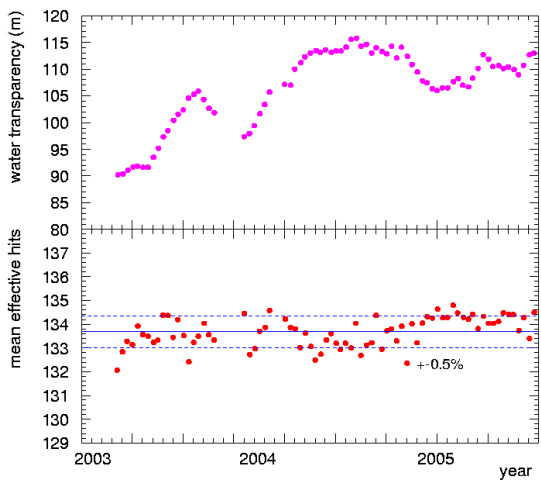}
\caption{Upper figure shows the time variation of the measured water transparency (weighed by the Cherenkov spectrum) during SK-II.  Lower figure shows the stability of the SK-II energy scale as a function of time.  The absence of data points in late 2003 is from detector dead time due to an electronics upgrade.}
\label{fig:water_transparency}
\end{center}
\end{figure}

Since the corrections in $N_{\mbox{\tiny{\textit{eff}}}}$ depend on the water transparency, the reconstructed energy also varies slightly with changing water quality.  See Figure~\ref{fig:water_transparency} for $N_{\mbox{\tiny{\textit{eff}}}}$ as a function of time for a given water transparency.  When calculating energy for data events, the water transparency value as determined by decay electrons from cosmic-ray muons is used as an input parameter.  However, for MC events, the change in water transparency is not simulated due to its relative stability compared to SK-I and a calculated, constant value of 101 m is used for all MC events in the analysis.

\begin{figure}[htbp]
\begin{center}
\includegraphics[scale=0.22]{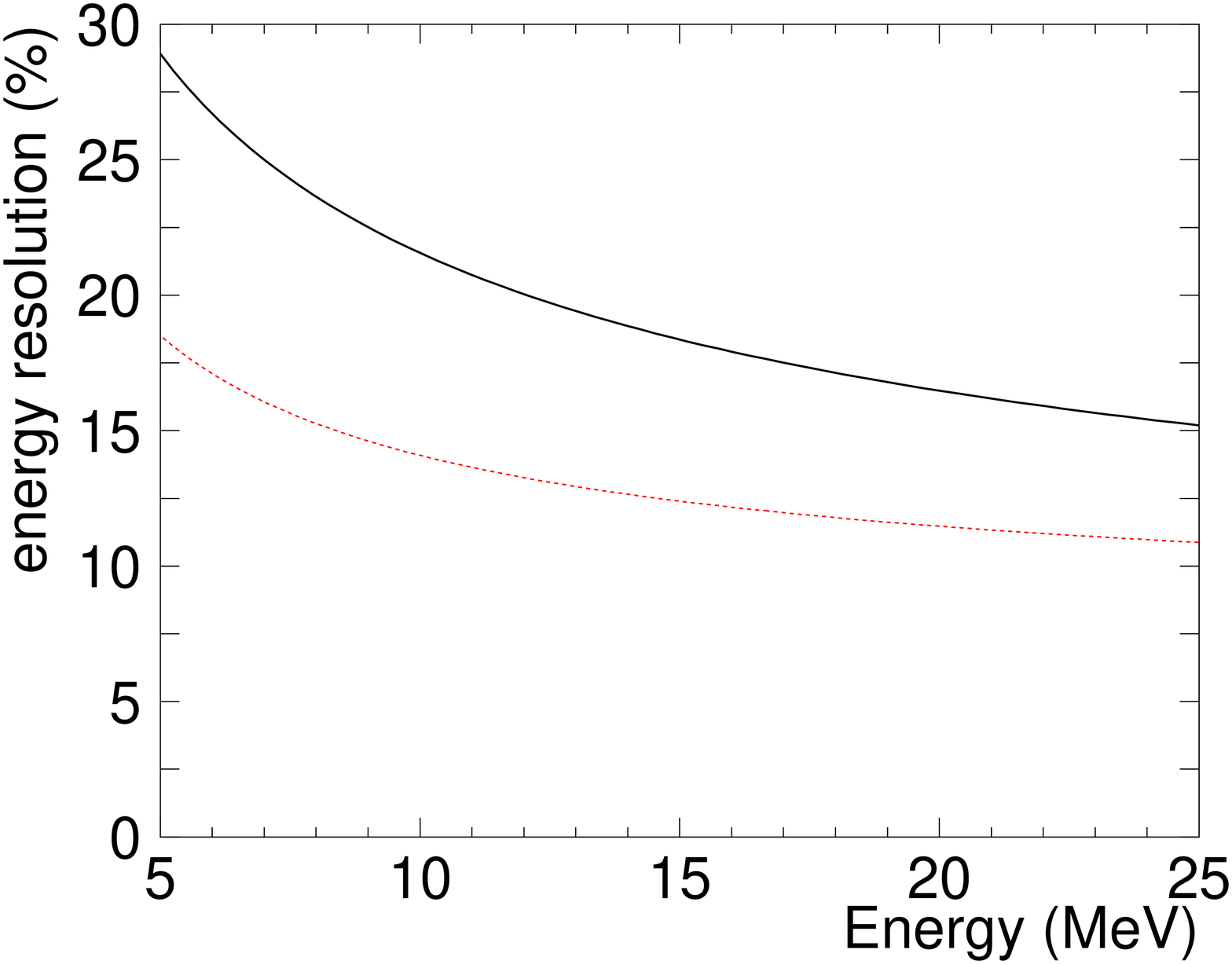}
\caption{Energy resolution as a function of total recoil electron energy of MC events.  The red dashed line is SK-I.}
\label{fig:eres_function}
\end{center}
\end{figure}

An analytical function of the detector's energy resolution can be determined with the same MC events used for the $N_{\mbox{\tiny{\textit{eff}}}}$-to-energy conversion function.  The energies of the MC events are calculated from the method described above and their fitted Gaussian mean energy and corresponding 1 sigma values are plotted for each discrete energy.  A sigma function $\sigma(E)$ is then fitted to use in a normal Gaussian probability density function
\begin{equation} \label{eq:eres1}
P(E,E')=\frac{1}{\sqrt{2 \pi} \sigma} \textrm{exp} \bigg[ -\frac{(E'-E)^2}{2 \sigma^2} \bigg],
\end{equation}
where $E$ is the electron's true recoil energy and $E'$ is the reconstructed energy.  The function $\sigma(E)$ for SK-II is given by
\begin{equation} \label{eq:eres2}
\sigma(E)=0.0536 + 0.5200 \sqrt{E} + 0.0458 E,
\end{equation}
in units of MeV.  The SK-I resolution is $\sigma=0.2468 + 0.1492 \sqrt{E} + 0.0690 E$.  Both resolutions are shown in Figure~\ref{fig:eres_function}.  Equation~\ref{eq:eres1} with Equation~\ref{eq:eres2} can be used to apply the SK resolution when calculating theoretical spectra for comparison with SK data.

\subsection{LINAC and $^{16}$N Energy Calibration}
\begin{figure}[htbp]
\begin{center}
\includegraphics[scale=0.45]{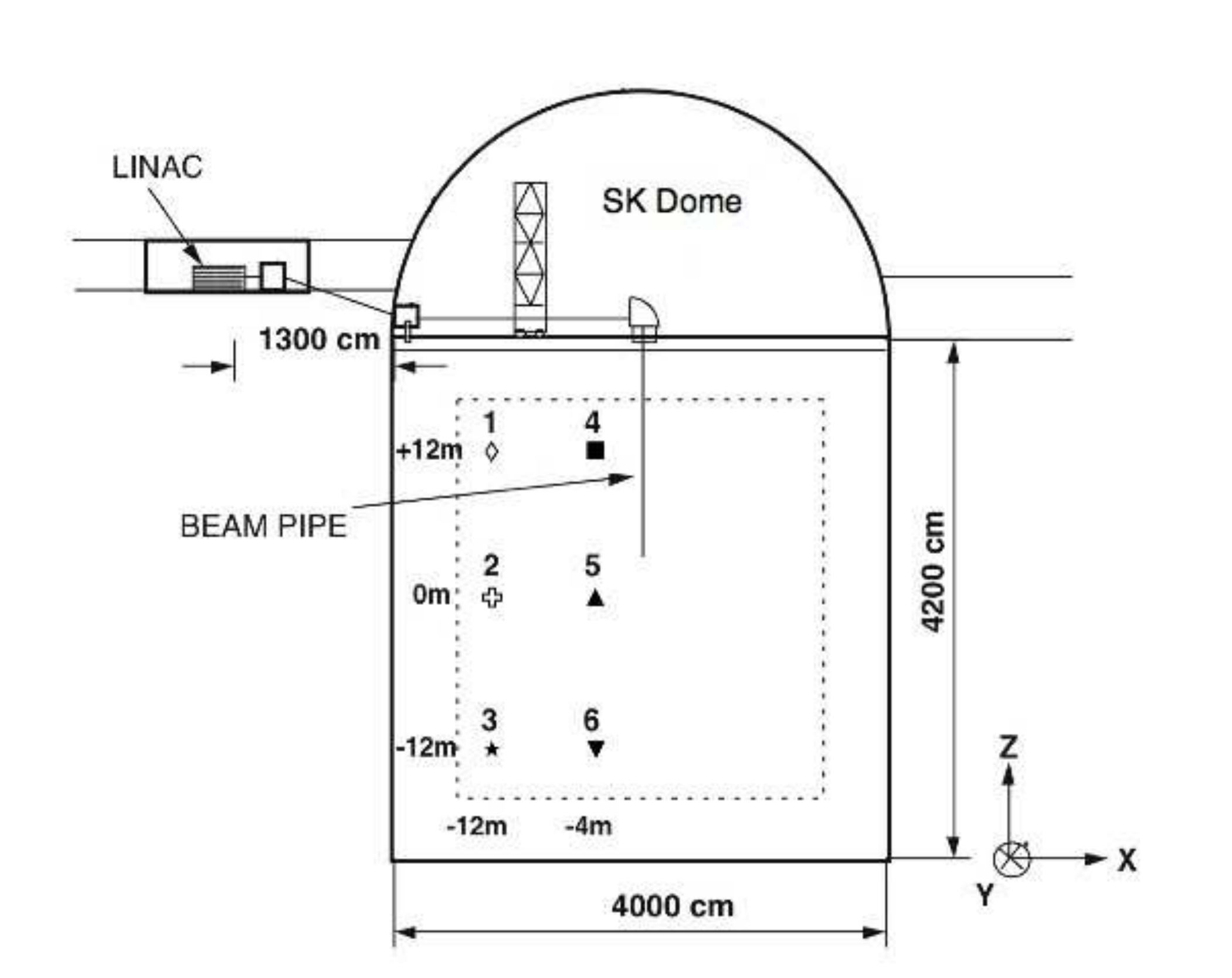}
\caption{The LINAC system at SK.  The dotted line represents the fiducial volume of the detector and the numbers 1-6 indicate where LINAC data were taken in SK-II.}
\label{fig:linac_diagram}
\end{center}
\end{figure}

As with SK-I, the primary instrument for energy calibration in SK-II is an electron linear accelerator [LINAC].  Detailed discussions on the LINAC calibration methods can be found elsewhere~\cite{linac}.  Electrons are injected into the SK tank at various positions (see Figure~\ref{fig:linac_diagram}) at energies between 5.8 and 13.4 MeV.  After reconstructing the energies of LINAC events, these data are compared with MC to determine the deviation in energy scales.  Various MC parameters are then adjusted to minimize the differences (see section II.A).

\begin{figure*}[htbp]
\begin{center}
\includegraphics[scale=0.38]{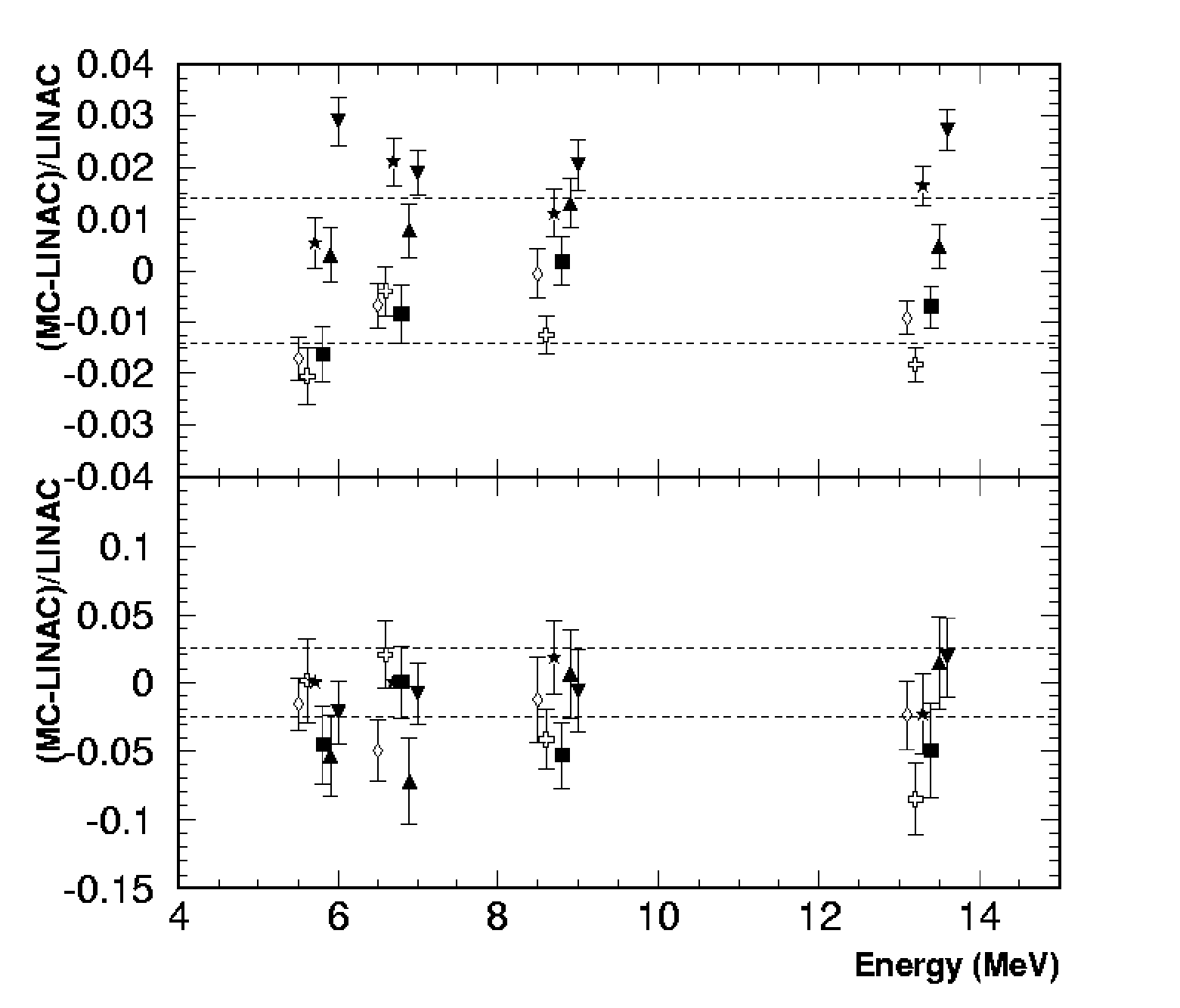}
\includegraphics[scale=0.4]{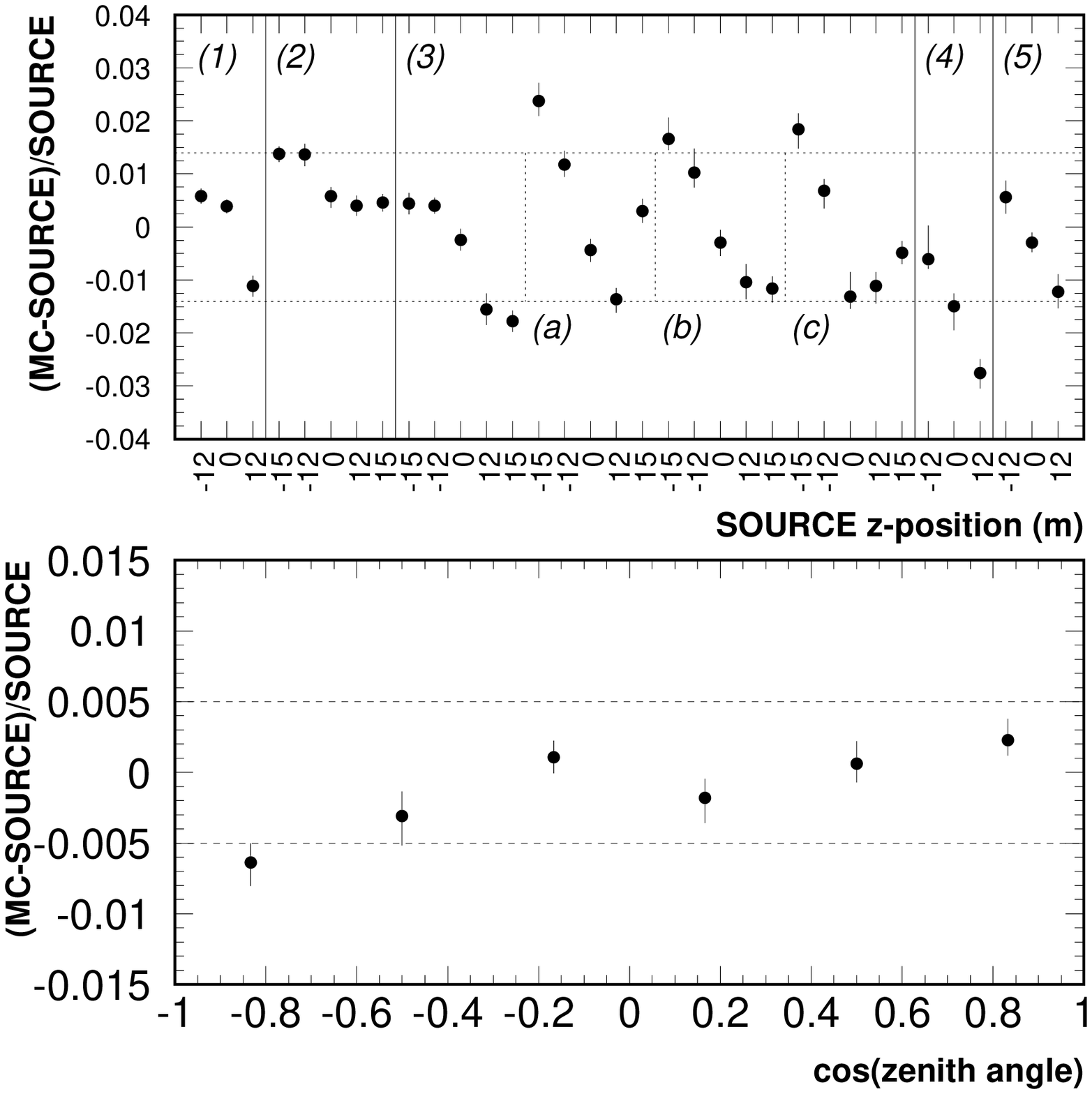}
\caption{Top left: Deviation in energy scale between LINAC MC and data.  Bottom left: differences in energy resolution between LINAC MC and data.  Refer to Figure~\ref{fig:linac_diagram} to relate data points with tank positions.  Top right: $^{16}$N energy scale deviation from MC.  Representing the varying times calibration data were taken, (1) Nov. 2003, (2) March 2004, (3) July 2004, (4) Nov. 2004, (5) Sep. 2005.  (a), (b), and (c) represent the x positions $15.20~\textrm{m}, 10.96~\textrm{m}, -14.49~\textrm{m}$ respectively.  All other calibration data were taken at $x=0.35~\textrm{m}$.  Bottom right: $^{16}$N energy scale deviation from MC for 6 selected zenith angles of the detector (-1 is down).}
\label{fig:linac_dtg}
\end{center}
\end{figure*}

The minimum uncertainty in the SK-II absolute energy scale using 13.4 and 8.8 MeV LINAC data is calculated to be 1.4\%.  This is in contrast to the SK-I estimated value of 0.64\%.  Figure~\ref{fig:linac_dtg} shows the relative difference of reconstructed energies of LINAC data and MC as well as their differences in energy resolution.

$^{16}$N is also used as a calibration source in conjunction with the LINAC calibration~\cite{dtg}.  $^{16}$N is produced by lowering a deuterium-tritium neutron generator into the tank and initiating the fusion reaction $^2$H$+ ^3$H$\to ^4$He$+ n$.  A fraction of these 14.2 MeV neutrons collide with $^{16}$O in the water to produce $^{16}$N which then decays with a half-life of 7.13 seconds.  In most cases, the Q-value is shared between 6.1 MeV gamma rays and a $\beta$-decay electron.

$^{16}$N decays allow directional studies on the energy scale not capable with the unidirectional LINAC beam.  At a total $^{16}$N decay product energy of 10.4 MeV, observed energy at several tank positions is compared with MC-simulated energy and the difference is shown to agree with those obtained from LINAC data and MC.  The $^{16}$N energy scale difference is averaged to be $\pm$1.2\% compared with LINAC's $\pm$1.4\%.  In addition, the isotropic $^{16}$N data are divided into zenith angle bins to show the relative asymmetry of the energy scale.  These show asymmetries within $\pm$0.5\% (Figure~\ref{fig:linac_dtg}) and are similar with SK-I values.

\begin{figure}[htbp]
\begin{center}
\includegraphics[scale=0.32, angle=0]{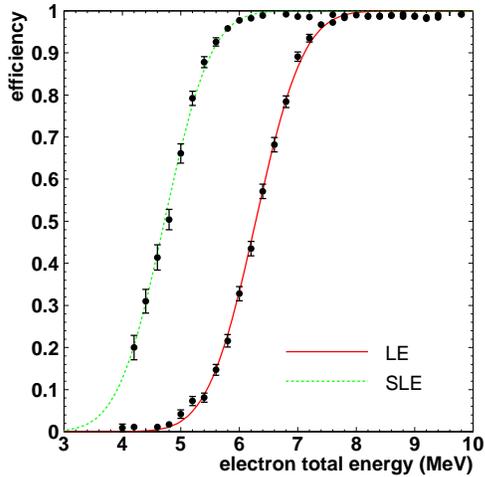}
\caption{Trigger efficiency as a function of energy.  The black dots are $^{16}$N calibration data and the lines are the best-fit error functions to the data (red solid is LE and green dashed is SLE).}
\label{fig:trigger_efficiency}
\end{center}
\end{figure}

Quantitative representations of trigger efficiencies are also obtained from $^{16}$N data.  The lowest threshold where the SK-II trigger is 100\% efficient is 6.5 MeV whereas the SK-I threshold is 4.5 MeV.  Figure~\ref{fig:trigger_efficiency} shows the trigger efficiencies for SK-II.  A systematic error is assigned to the trigger efficiency by comparing the value given by $^{16}$N data and MC-simulated trigger events (0.5\% on the total flux measurement).

\section{Data Analysis}
\subsection{Trigger Scheme}
Like SK-I, SK-II has two levels of triggering for solar neutrino analysis: low energy [LE] and super-low energy [SLE] thresholds which require a minimum of about 14 and 10 hit PMTs respectively to register an event.  At the beginning of data taking in December 2002, only the LE trigger threshold was applied.  At and above 8.0 MeV is where the LE trigger is 100\% efficient.  Later, the threshold was lowered and SLE data was taken with 100\% efficiency at 6.5 MeV.  Ultimately, the LE+SLE analysis threshold was set to 7.0 MeV due to the large number of background events below this level.  The LE analysis period lasted from December 24th, 2002 to July 14th, 2003 for an exposure of 159 live days.  The LE+SLE period lasted from July 15th, 2003 until October 5th, 2005 for an exposure of 632 live days.

SLE triggered events are filtered online to reduce the amount of data written to limited storage space.  Events reconstructed outside the fiducial volume are rejected.  The data are reduced by a factor of approximately six.  See Fig~\ref{fig:bonsai_resolution} for the vertex resolution as a function of energy of the online fitter.

\subsection{Background Reduction}
For SK-II, we implement a new two-part cut of events based on defined goodness functions of PMT timing and hit patterns.

Many background events remain due to mis-reconstruction after the usual two-meter fiducial volume cut, (which reduces background coming from the PMTs and blast shields.)  Whereas a gamma-ray cut solely relies on vertex and directional reconstruction, the 2-dimensional timing-pattern cut removes those events whose reconstruction should not be trusted.  An optimized hit PMT timing goodness is defined (Equation~\ref{eq:goodness}) by comparing two timing residual Gaussian distributions, one with a width of $\sigma=$ 5 ns to encompass selected hits and the other with a $\omega=$ 60 ns width characteristic of the PMT timing resolution for a single photo-electron:

\begin{equation} \label{eq:goodness}
g_t(\vec{v})=\frac{\Sigma e^{-\frac{1}{2}\big((\frac{\tau_i(\vec{v})-t_0}{\omega}) + (\frac{\tau_i(\vec{v})-t_0}{\sigma})\big)^2}}{\Sigma e^{-\frac{1}{2}(\frac{\tau_i(\vec{v})-t_0}{\omega})^2}}.
\end{equation}
The effective hit time is defined as $\tau_i(\vec{v})=t_i - |\vec{v} - \vec{h}_i|/c$ which is just the timing residual $t_{residual}$ of Equation~\ref{eq:t_residual} with added $t_0$.  The sums are over all hits.

The hit pattern goodness allows us to identify Cherenkov events by their azimuthal-symmetric ring pattern from a reconstructed vertex and direction.  All others are labeled mis-reconstructed or non-Cherenkov events.  A goodness function $g_p(\vec{v})$ is defined for all directional events.

\begin{figure}[htbp]
\begin{center}
\includegraphics[scale=0.24]{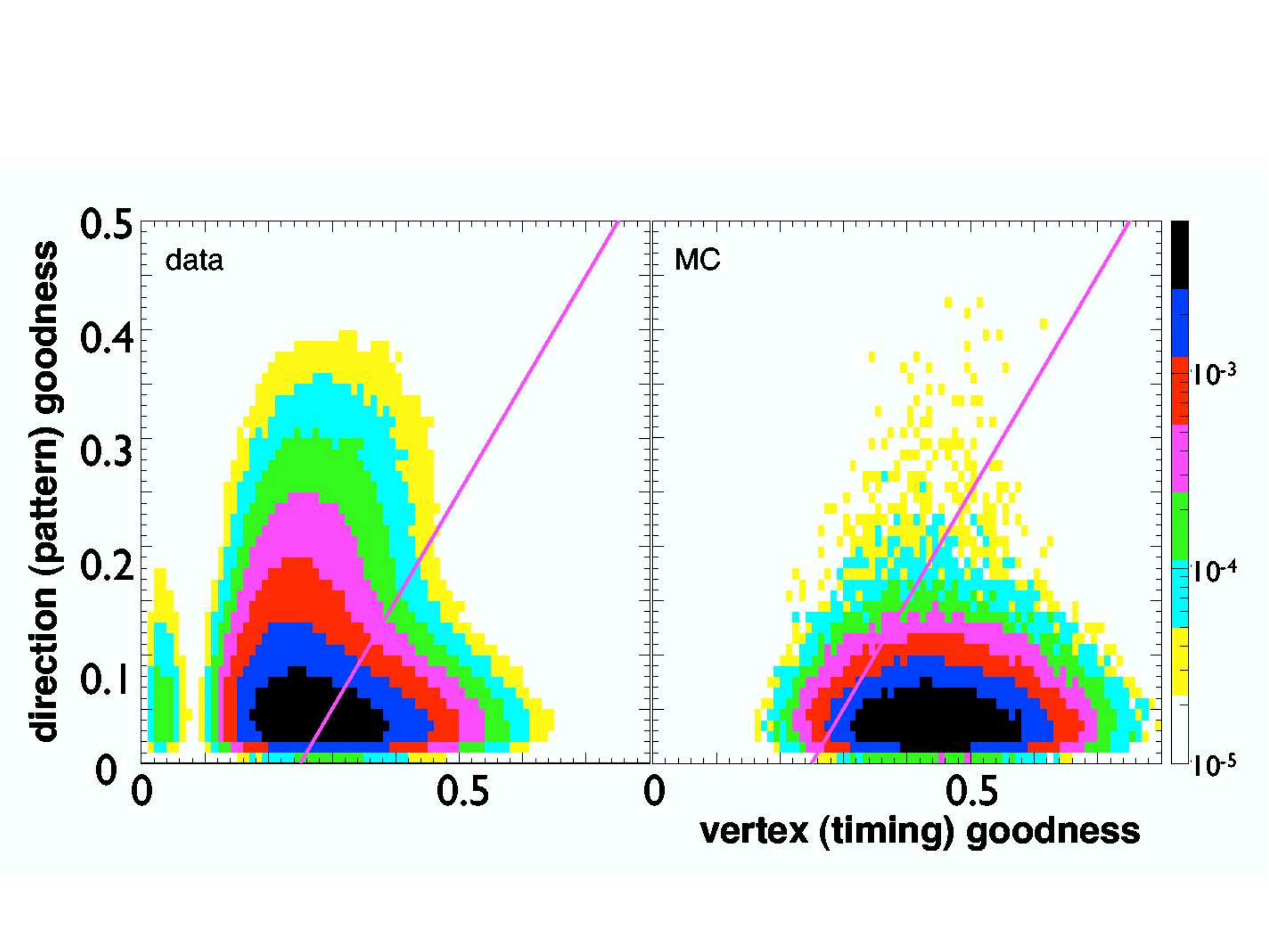}
\caption{PMT timing and hit pattern cut.  Data (left) show an excess of mis-reconstructed and non-Chrerenkov events to the upper-left of the diagonal cut line.  Approximately 78\% (8\%) of data (MC) events between 7.0-7.5 MeV are rejected by the cut.  The color scale is to show the relative (normalized) number of events.}
\label{fig:dirks}
\end{center}
\end{figure}

A cut on the goodness values is made in tandem using a hyperbolic radius of $g_t^2 - g_p^2 > 0.25$ and rejecting all other events.  Figure~\ref{fig:dirks} shows this background reduction cut on data and $^8$B Monte Carlo in the 7.0-7.5 MeV bin.  When the cut is applied to LINAC data and MC, a total flux systematic error of $\pm$3.0\% is obtained.

\begin{figure}[htbp]
\begin{center}
\includegraphics[scale=0.24]{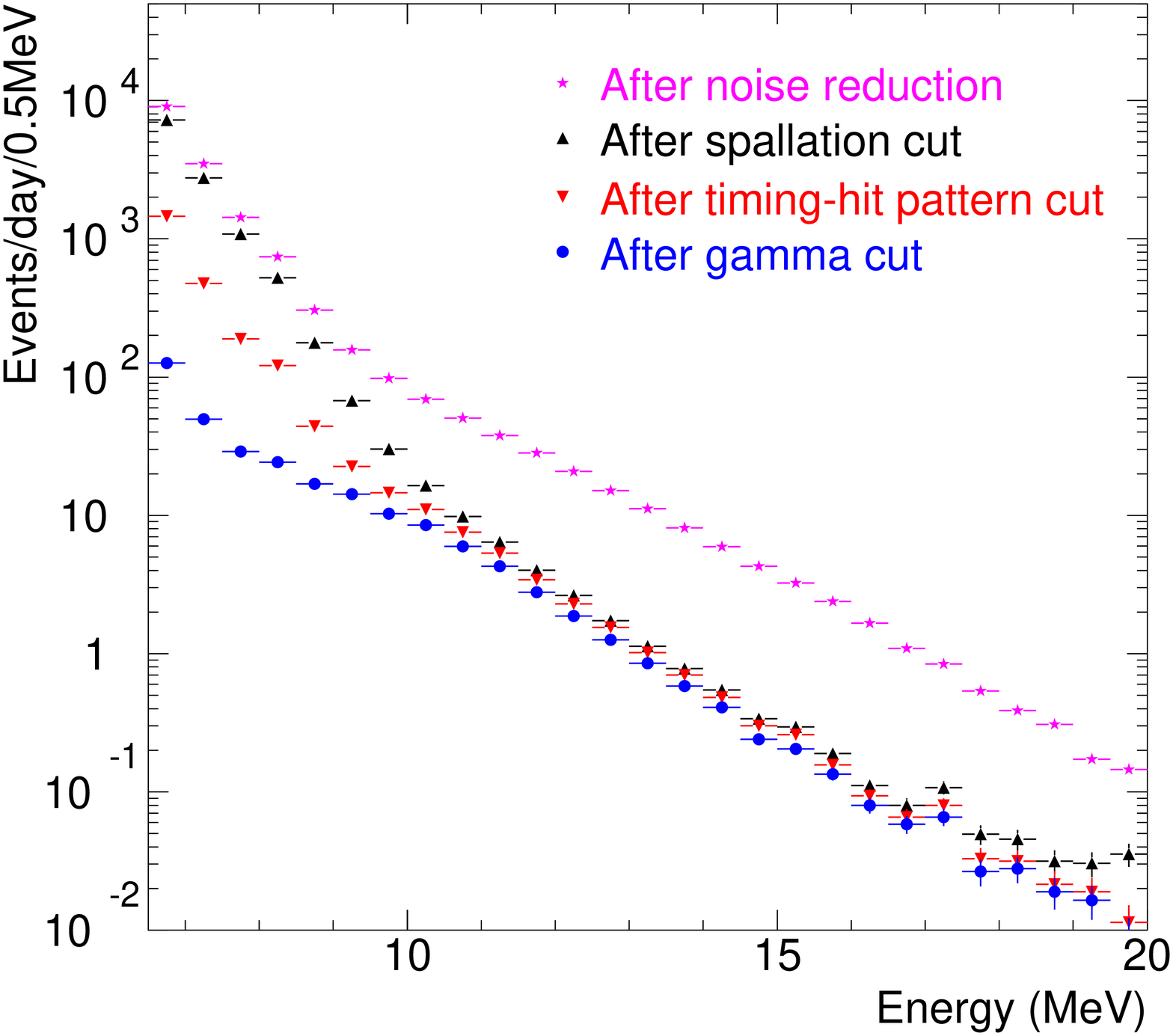}
\includegraphics[scale=0.35]{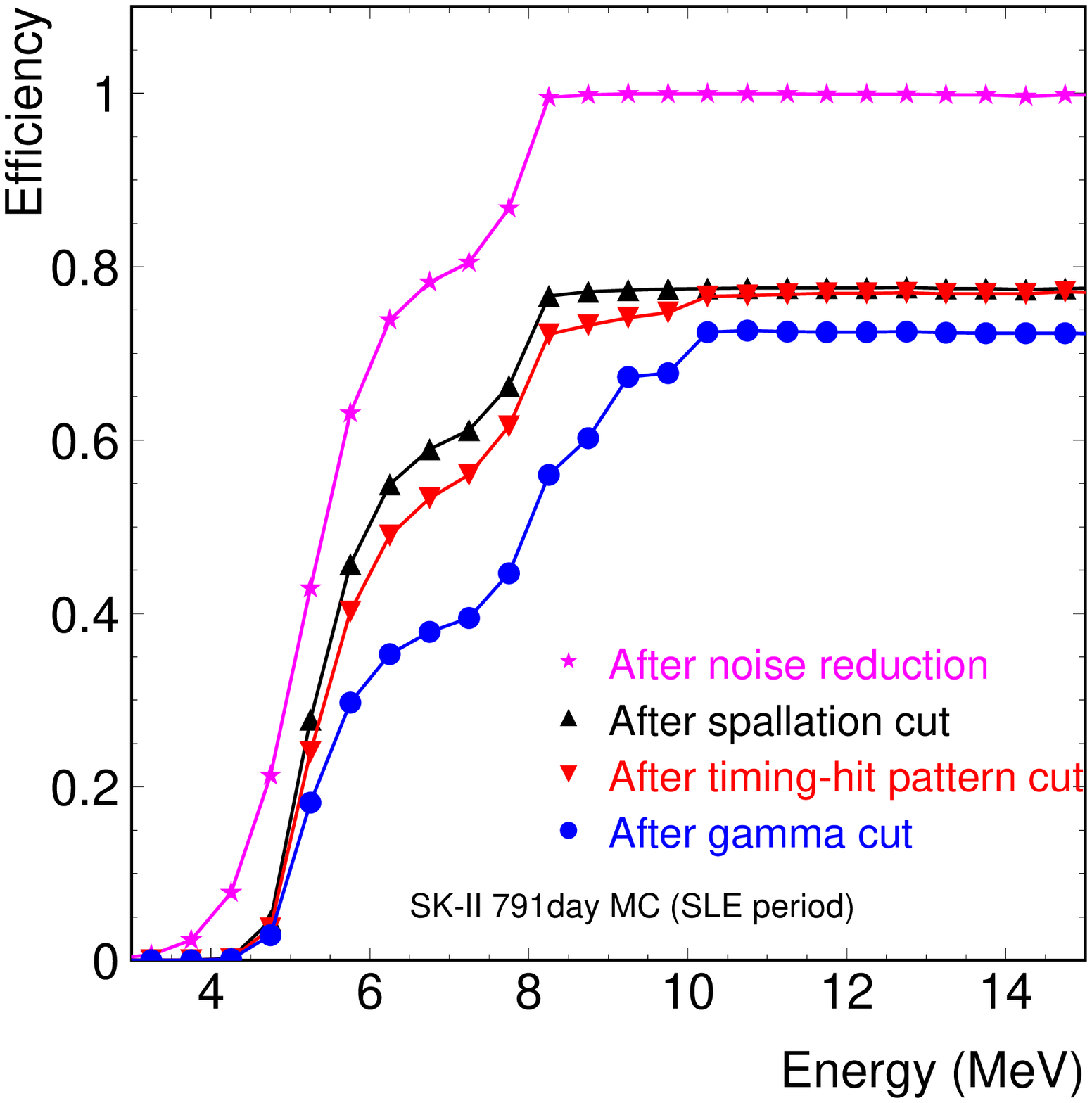}
\caption{Summary of the data reduction steps (top) and their efficiencies on MC (bottom)}
\label{fig:background_reduction}
\end{center}
\end{figure}

The timing-hit pattern cut is treated as a second reduction after the removal of noise and spallation events from the initial data set.  All cuts and their efficiencies are shown in Figure~\ref{fig:background_reduction}.

\begin{table}[htbp]  \label{tab:systematic_errors}
\begin{center}
 \caption{SK-II systematic error of each item in $\%$.  Numbers in parentheses are the values obtained from calibration data before application to the neutrino flux.}
  \begin{tabular}{l|c|c}
   \hline
   \hline
   & flux & day-night \\
   \hline
   \hline
Energy scale (absolute $\pm 1.4 \%$) & $+4.2-3.9$ & \\
Energy scale (relative $\pm 0.5 \%$) & & $\pm$1.5 \\
Energy resolution (2.5 $\%$) & $\pm$0.3 & \\
$^8$B spectrum & $\pm$1.9 & \\
Trigger efficiency & $\pm$0.5 & \\
1$^{st}$ reduction & $\pm$1.0 & \\
2$^{nd}$ reduction & $\pm$3.0 & \\
Spallation dead time & $\pm$0.4 & \\
Gamma cut & $\pm$1.0 & \\
Vertex shift & $\pm$1.1 & \\
Non-flat background & $\pm$0.4 & $\pm$3.4 \\
Angular resolution (6.0$\%$) & $\pm$3.0 & \\
Cross section & $\pm$0.5 & \\
Live time & $\pm$0.1 & $\pm$0.1 \\
\hline
Total & $+6.7-6.4$ & $\pm$3.7 \\
 \hline
  \hline
  \end{tabular}
 \end{center}
\end{table}

\subsection{Total Flux Result}
\begin{figure}[htbp]
\begin{center}
\includegraphics[scale=0.11]{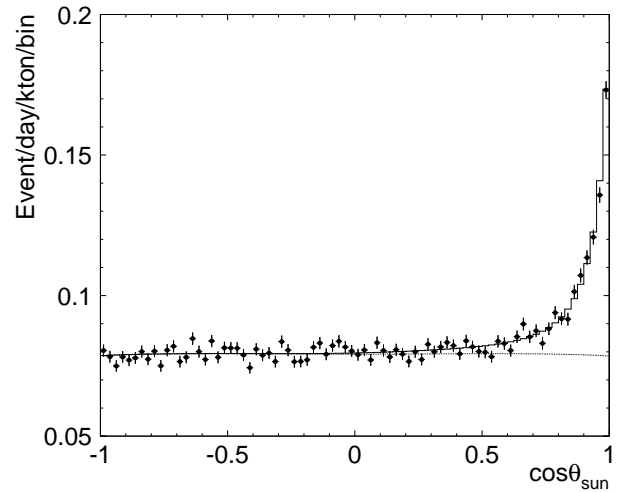}
\caption{The angular distribution of solar neutrino candidate events.  The flat line seen under the peak in the solar direction represents background contributions.}
\label{fig:solar_peak}
\end{center}
\end{figure}

The SK-II solar neutrino signal is extracted from the strongly forward biased direction of recoil electrons from $\nu$-e elastic scattering.  A likelihood fit to the signal and background is utilized to determine the flux.  For a live time of 791 days of SK-II data from 7.0 to 20.0 MeV, the extracted number of signal events is 7212.8$^{+152.9}_{-150.9}$(stat.) $^{+483.3}_{-461.6}$(sys.).  The corresponding $^8$B flux is:
\begin{eqnarray*}
(2.38 \pm 0.05 (\textrm{stat.}) ^{+0.16}_{-0.15} (\textrm{sys.})) \times 10^ 6~\textrm{cm}^{-2} \textrm{sec}^{-1}.
\end{eqnarray*}
It is statistically consistent with the SK-I value of $(2.35 \pm 0.02 (\textrm{stat.}) \pm 0.08 (\textrm{sys.})) \times 10^ 6~\textrm{cm}^{-2} \textrm{sec}^{-1}$.  The systematic uncertainties of SK-I and II are mostly uncorrelated due to differences in energy scale, event selection, event reconstruction methods, etc.  Figure~\ref{fig:solar_peak} shows the angular distribution of extracted solar neutrino events.  Table I lists the SK-II systematic errors assigned for the total flux and day-night difference.

\subsection{Time Variation Results}
\subsubsection{Day-Night and Seasonal Variation}
Time variations of the solar neutrino flux are also determined by looking at day and night fluxes and the change in total flux at regular intervals during the live time of SK-II.  The day and night fluxes are measured by selecting events which occur when the cosine of the solar zenith angle is less than zero (day) and greater than zero (night).  Unlike the total flux, the day and night fluxes are quoted using a threshold of 7.5 MeV due to low signal to noise ratio for the 7.0-7.5 MeV bin in the solar direction after the data set is divided.  Their values are
\begin{eqnarray*}
\Phi_{day} & = & (2.31 \pm 0.07 (\textrm{stat.}) \pm 0.15 (\textrm{sys.})) \times 10^6~\textrm{cm}^{-2} \textrm{sec}^{-1},
\vspace{0.05cm}\\
\Phi_{night} & = & (2.46 \pm 0.07 (\textrm{stat.}) \pm 0.16 (\textrm{sys.})) \times 10^6~\textrm{cm}^{-2} \textrm{sec}^{-1}.
\end{eqnarray*}
With these fluxes, the asymmetry value is found from ${\cal A} =(\Phi_{day} - \Phi_{night})/( \frac{1}{2} (\Phi_{day} + \Phi_{night}))$.  The SK-II day-night difference yields
\begin{eqnarray*}
{\cal A} = -0.063 \pm 0.042 (\textrm{stat.}) \pm 0.037 (\textrm{sys.}).
\end{eqnarray*}
As with SK-I (${\cal A} = -0.021 \pm 0.020 (\textrm{stat.})^{+0.013}_{-0.012} (\textrm{sys.})$), no day-night asymmetry is discerned from the SK-II solar data set.  The SK-I asymmetry value is statistically consistent with SK-II.

\begin{figure}[htbp]
\begin{center}
\includegraphics[scale=0.2475]{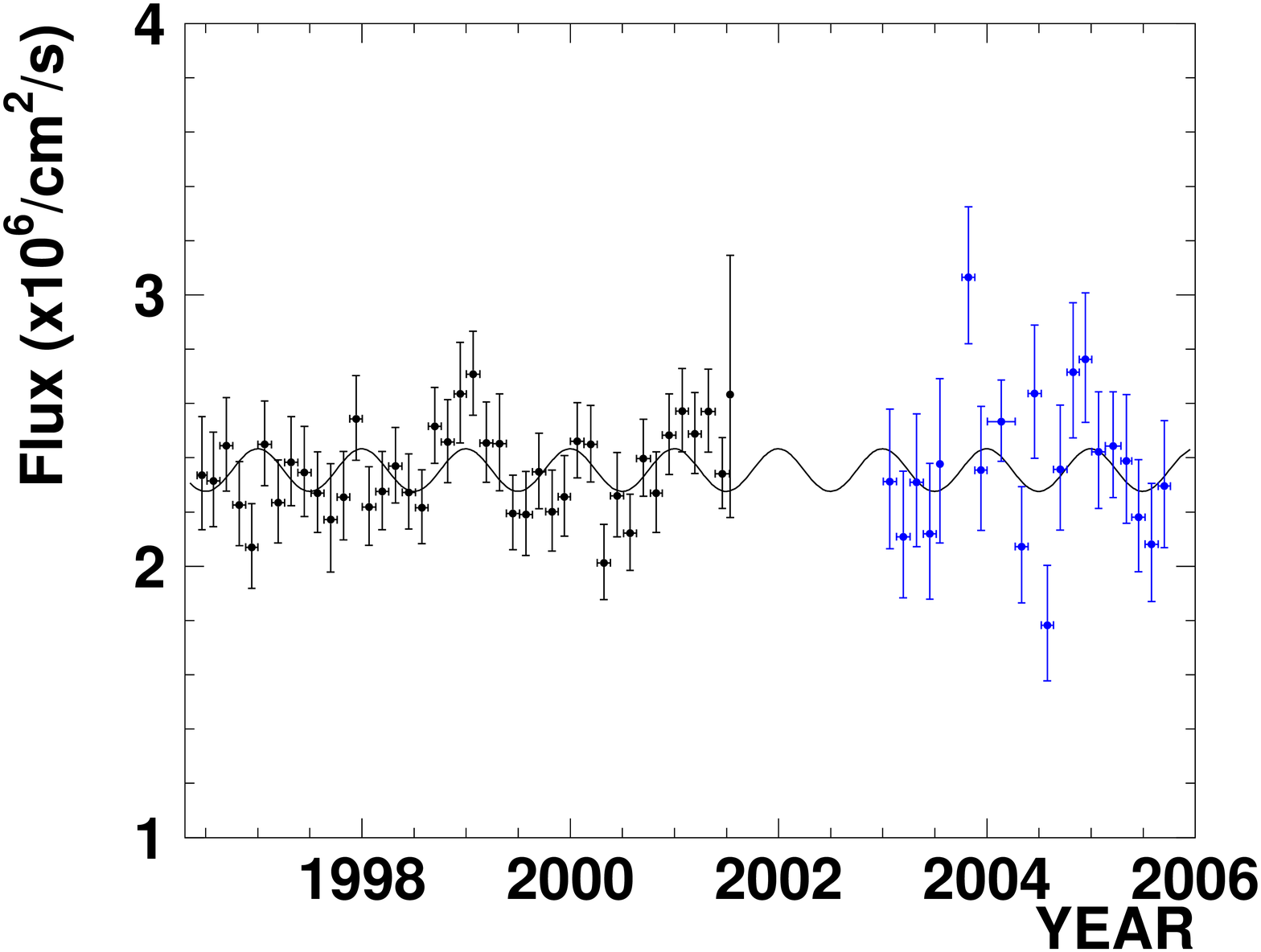}
\caption{Time dependence of the solar neutrino flux.  The black points are from the 1496-day SK-I data set at a threshold of 5.0 MeV.  The blue points are from the 791-day SK-II data set at a threshold of 7.0 MeV.  The black line represents the expected $1/r^2$ flux variations due to the eccentricity of the earth's orbit around the sun.  Errors are statistical only.  The absence of data points between SK-I and SK-II indicates dead time while construction of SK-II was occurring.}
\label{fig:time_variation}
\end{center}
\end{figure}

The total flux variation as a function of time, or seasonal variation, for both SK-I and SK-II solar data is shown in Figure~\ref{fig:time_variation}.  Each bin represents 1.5 months and is seen to follow a sinusoidal trend consistent with the expected $1/r^2$ flux variations due to the eccentricity of the earth's orbit around the sun.  SK-II has excellent agreement with SK-I data, thus showing the continuation of the SK solar neutrino measurement through two phases of the detector.

\subsubsection{Flux Correlation with Solar Activity}
With the completion of SK-II, the solar neutrino flux measurement of the Super-Kamiokande experiment spans an interval of 9.5 years.  This closely coincides with the full period of solar cycle 23.  To address any possible correlation of solar neutrino flux with sun spot number, the SK-I and II flux time variation data are compiled in 1-year bins between 1996 and 2006.  The SK-I data set (from 1996 to 2001) is taken from a 5.0 MeV threshold while SK-II is from 7.0 MeV.  Errors are statistical only.  From 1996 to the end of the SK-II phase in October 2005, the solar neutrino flux is stable and shows no pattern of correlation with the minima and maximum of solar cycle 23.  This is consistent with (and a continuation of) the Kamiokande measurement and comparison with solar cycle 22~\cite{kamiokande_solar_active}, albeit with a greater level of precision for Super-Kamiokande.

\begin{figure}[htbp]
\begin{center}
\includegraphics[scale=0.2475]{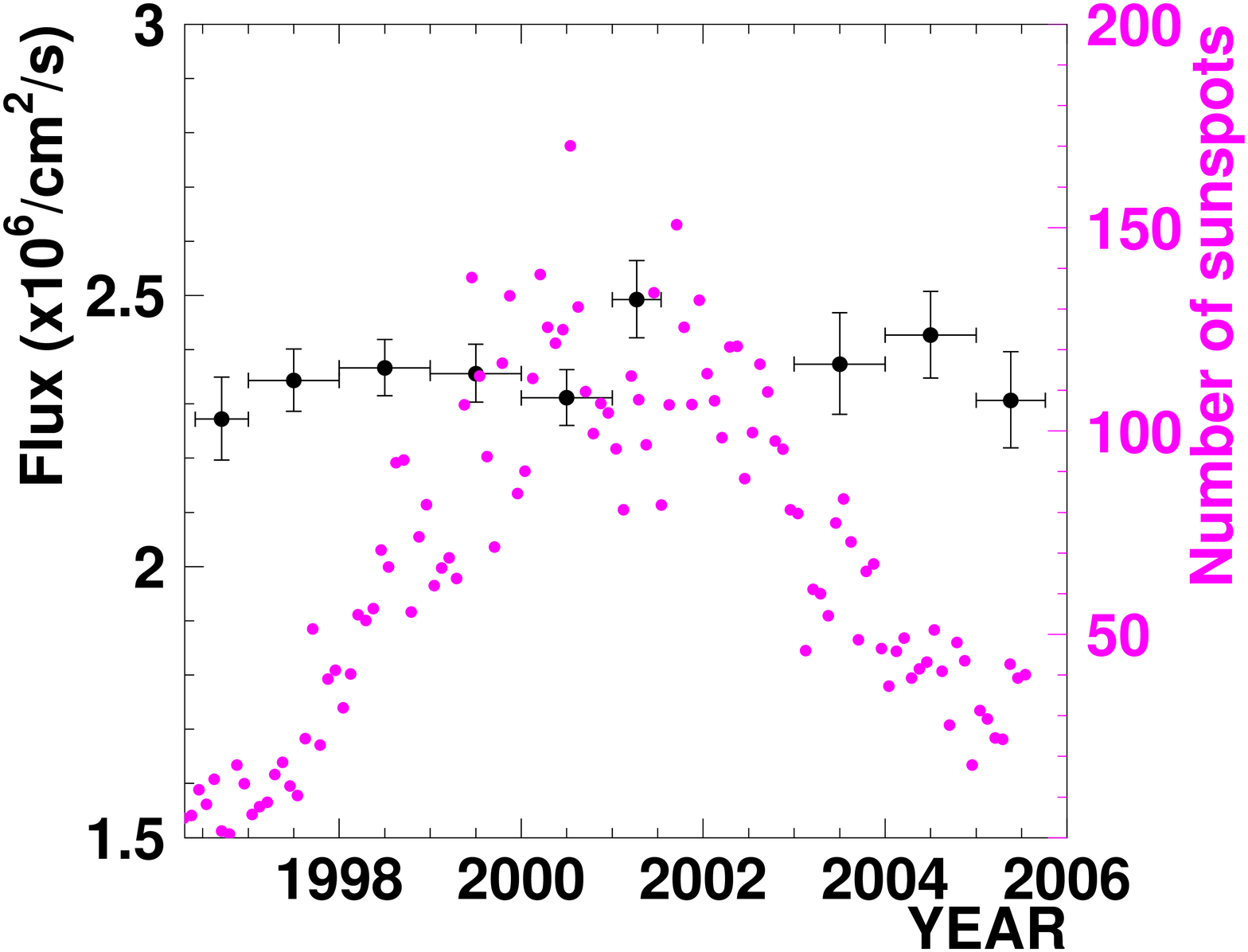}
\caption{Time variation of the solar neutrino flux overlaid with sun spot number for solar cycle 23.  Errors are statistical only.  The SK-I and II 1-year binned solar flux data gives an agreement of $\chi^2=6.11$ (52\% c.l.) when compared to a straight line.}
\label{fig:solar_active}
\end{center}
\end{figure}

\subsection{Energy Spectrum}
\begin{figure}[htbp]
\begin{center}
\includegraphics[scale=0.2475]{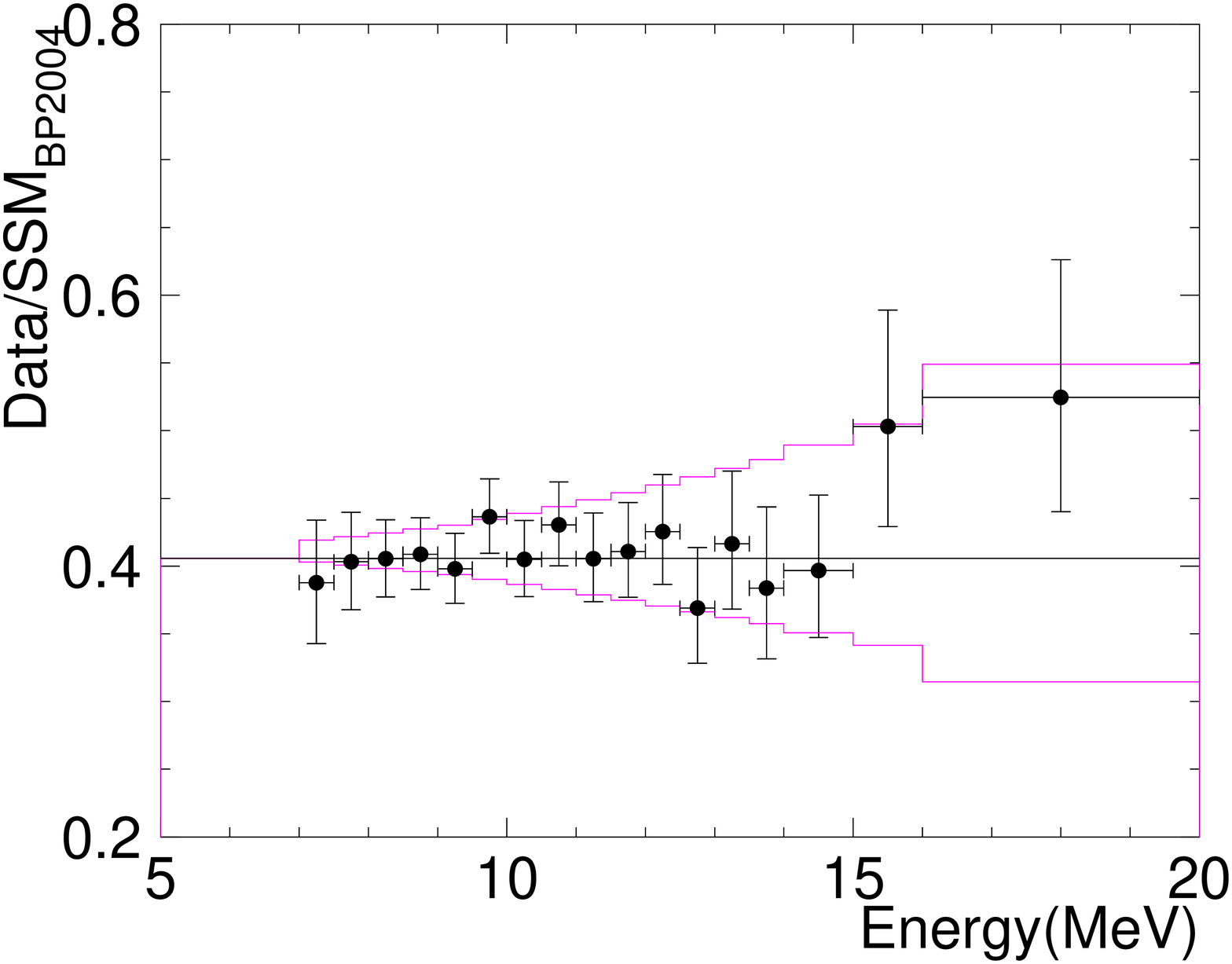}
\caption{Ratio of observed and expected energy spectra.  The purple lines represent a $\pm1$ sigma level of the energy correlated systematic errors.  The black line represents the SK-I 1496-day average and shows agreement with SK-II.}
\label{fig:energy_spectrum}
\end{center}
\end{figure}

The recoil electron energy spectrum is obtained by dividing the total flux into 17 energy bins ranging from 7.0 to 20.0 MeV.  The bin boundaries and flux values are listed in Table II.  Figure~\ref{fig:energy_spectrum} shows the observed energy spectrum divided by the expected spectrum without oscillation determined from the BP2004 SSM~\cite{ssm}.  The line through the spectrum represents the total SK-I 1496-day average.  Again, as with the seasonal variation, SK-II shows excellent agreement with SK-I.

\begin{table*}[htbp]   \label{tab:rates}
 \begin{center}
\caption{SK-II observed energy spectra expressed in units of event/kton/year.  The errors in the observed rates are statistical only.  The 7.0-7.5 MeV energy bin is excluded from the day-night analysis.  Correction is made for the reduction efficiencies in Figure~\ref{fig:background_reduction}.  The expected rates neglecting oscillation are for the BP2004 SSM flux values.  $\theta_{z}$ is the angle between the z-axis of the detector and the vector from the Sun to the detector.}
  \begin{tabular}{c|c|c|c|c|c}
   \hline
   \hline
   Energy    & \multicolumn{3}{c}{Observed rate} & \multicolumn{2}{c}{Expected rate}\\
   (MeV)     &  ALL & DAY & NIGHT & $^8$B & hep\\
             & $ -1 \leq \cos\theta_{\rm z} \leq 1 $ 
             & $ -1 \leq \cos\theta_{\rm z} \leq 0 $ 
             & $  0 <    \cos\theta_{\rm z} \leq 1 $  & &\\ 
   \hline
   \hline

 $ 7.0- 7.5$ & $ 43.7^{+  5.2}_{-  5.1}$ & $-$ & $-$ & 112.4 & 0.257 \\ 
 $ 7.5- 8.0$ & $ 40.0^{+  3.6}_{-  3.5}$ & $ 36.4^{+  5.1}_{-  4.9}$ & $ 43.6^{+  5.2}_{-  5.0}$ &  99.1 & 0.245 \\ 
 $ 8.0- 8.5$ & $ 34.9^{+  2.5}_{-  2.4}$ & $ 34.4^{+  3.5}_{-  3.4}$ & $ 35.5^{+  3.5}_{-  3.4}$ &  85.9 & 0.231 \\ 
 $ 8.5- 9.0$ & $ 30.1^{+  2.0}_{-  1.9}$ & $ 27.0^{+  2.8}_{-  2.7}$ & $ 33.0^{+  2.8}_{-  2.7}$ &  73.5 & 0.215 \\ 
 $ 9.0- 9.5$ & $ 24.5^{+  1.6}_{-  1.6}$ & $ 23.9^{+  2.3}_{-  2.2}$ & $ 25.0^{+  2.3}_{-  2.2}$ &  61.4 & 0.198 \\ 
 $ 9.5-10.0$ & $ 22.0^{+  1.4}_{-  1.4}$ & $ 20.7^{+  2.0}_{-  1.9}$ & $ 23.3^{+  2.0}_{-  1.9}$ &  50.3 & 0.181 \\ 
 $10.0-10.5$ & $ 16.6^{+  1.2}_{-  1.1}$ & $ 15.4^{+  1.7}_{-  1.6}$ & $ 17.6^{+  1.7}_{-  1.6}$ &  40.7 & 0.163 \\ 
 $10.5-11.0$ & $ 13.9^{+  1.0}_{-  1.0}$ & $ 13.5^{+  1.5}_{-  1.4}$ & $ 14.2^{+  1.5}_{-  1.4}$ &  32.1 & 0.145 \\ 
 $11.0-11.5$ & $ 10.3^{+  0.9}_{-  0.8}$ & $ 11.3^{+  1.3}_{-  1.2}$ & $  9.4^{+  1.2}_{-  1.1}$ &  25.3 & 0.129 \\ 
 $11.5-12.0$ & $8.06^{+ 0.71}_{- 0.66}$ & $ 7.11^{+ 1.00}_{- 0.90}$ & $ 8.96^{+ 1.03}_{- 0.94}$ & 19.51 & 0.113 \\ 
 $12.0-12.5$ & $6.28^{+ 0.62}_{- 0.58}$ & $ 6.82^{+ 0.94}_{- 0.84}$ & $ 5.79^{+ 0.86}_{- 0.77}$ & 14.67 & 0.098 \\ 
 $12.5-13.0$ & $4.07^{+ 0.50}_{- 0.45}$ & $ 4.18^{+ 0.73}_{- 0.63}$ & $ 3.97^{+ 0.70}_{- 0.61}$ & 10.96 & 0.084 \\ 
 $13.0-13.5$ & $3.32^{+ 0.43}_{- 0.38}$ & $ 2.95^{+ 0.62}_{- 0.53}$ & $ 3.66^{+ 0.61}_{- 0.53}$ &  7.91 & 0.071 \\ 
 $13.5-14.0$ & $2.23^{+ 0.35}_{- 0.30}$ & $ 2.95^{+ 0.57}_{- 0.48}$ & $ 1.59^{+ 0.44}_{- 0.35}$ &  5.74 & 0.060 \\ 
 $14.0-15.0$ & $2.77^{+ 0.39}_{- 0.35}$ & $ 2.99^{+ 0.60}_{- 0.51}$ & $ 2.58^{+ 0.53}_{- 0.45}$ &  6.90 & 0.091 \\ 
 $15.0-16.0$ & $1.75^{+ 0.30}_{- 0.26}$ & $ 1.37^{+ 0.42}_{- 0.32}$ & $ 2.08^{+ 0.45}_{- 0.37}$ &  3.41 & 0.063 \\ 
 $16.0-20.0$ & $1.37^{+ 0.27}_{- 0.22}$ & $ 1.11^{+ 0.37}_{- 0.28}$ & $ 1.60^{+ 0.40}_{- 0.31}$ &  2.52 & 0.089 \\ 

 \hline
  \hline
  \end{tabular}
 \end{center}
\end{table*}

\section{SK-II Oscillation Analysis}
\subsection{$\chi^2$ Minimization}
Oscillations of solar neutrinos have been studied by numerous experiments and have placed increasingly stringent constraints on the mixing angle between neutrino mass and flavor eigenstates as well as neutrino mass difference.  In the statistically large data sample of SK-I, those constraints, assuming two flavor oscillations, favor the large mixing angle [LMA] region at 95\% confidence level.  The best fit values are given in the LMA region at $\tan^2\theta$=0.52 and $\Delta m^2=6.3\times10^{-5}\textrm{eV}^2$.  The favored regions and corresponding best fit value are from a fit to the SK-I spectrum and time variation rates.  The $^8$B flux is also constrained by the SNO total rate~\cite{snosalt}.

The determination of the solar neutrino oscillation parameters ($\theta_{12}, \Delta m_{12}$) in SK-II is accomplished in much the same way as the previous SK-I result.  Two neutrino oscillation is assumed and for each set of oscillation parameters, a $\chi^2$ minimization of the total $^8$B and hep neutrino flux is fit to the data.  The entire SK-II observed spectrum is utilized from a 7.0 MeV threshold.  The expected oscillated $^8$B and hep flux is derived from numerically calculated MSW~\cite{msw} $\nu_e$ survival probabilities and the unoscillated flux provided by the SSM.  It is then converted to an expected SK-II rate spectrum by utilizing the $\nu - e$ elastic scattering cross section and the SK-II detector's energy resolution.  To account for the systematic uncertainties in energy resolution as well as the energy scale and the $^8$B neutrino spectrum model shape, the combined rate predictions are modified by energy shape factors, $f(E_i,\delta_B, \delta_S, \delta_R)$.  $\delta_B$, $\delta_S$, and $\delta_R$ represent unfcertainty in the $^8$B neutrino spectrum, SK-II energy scale, and SK-II energy resolution respectively.  The function $f$ serves to shift the rate predictions corresponding to a given uncertainty $\delta$ in the data rate.  The following equation shows the SK-II spectrum $\chi^{2}$ along with energy correlated systematic error shape factors applied to the expected rate:
\begin{displaymath}
\chi^2_{\textrm{SK-II}} = \sum_{i=1}^{17} \frac{(d_i - (\beta b_i + \eta h_i) \times f(E_i,\delta_B,\delta_S,\delta_R))^2}{\sigma_i^2}+
\end{displaymath}
\begin{equation}
\left(\frac{\delta_B}{\sigma_B} \right)^2
+ \left( \frac{\delta_S}{\sigma_S} \right)^2
+ \left( \frac{\delta_R}{\sigma_R} \right)^2
+2 \Delta \log(\cal{L}),
\end{equation}
where $d_i$ is the observed rate divided by the expected, unoscillated rate for the $i^{th}$ energy bin.  Similarly, $b_i$ and $h_i$ are the predicted MSW oscillated rates divided by the unoscillated rate for $^8$B and hep neutrinos respectively.  $\beta$ ($\eta$) scales the $^8$B (hep) neutrino flux.  The last term is the unbinned time-variation likelihood to the SK-II solar zenith angle flux variation above a 7.5 MeV threshold.  This likelihood is analogous to the one used in SK-I.

The energy uncorrelated systematic uncertainty is assigned to 4.8\% (the quadrature sum of the energy independent errors in Table I) and is used to describe the error on the total rate.  For the spectrum rate uncertainties, the value 4.8\% is conservatively assigned to each bin and is added in quadrature to the statistical error (Table II) to equal $\sigma_i$ in the SK-II $\chi^2$.  See the appendix of~\cite{unique_solution} for more details.

\subsection{Oscillation Results - SK Only}
A minimization of the $\chi^2$ in the previous section yields excluded regions when $\beta$ and $\eta$ are left unconstrained.  By constraining the $^8$B flux to the total NC flux value from SNO~\cite{snosalt}, allowed parameter regions can be obtained.  Figure~\ref{fig:contours} shows both excluded and allowed regions at 95\% confidence level.  They are consistent with previous SK-I results.  The primary constraint in SK-II is from the time-variation data although some spectral exclusion is also seen at $\Delta m^2 \approx 10^{-4} \textrm{eV}^2$.  The same oscillation analysis is performed while including $\chi^2$ terms corresponding to the SK-I values (namely, the spectrum and unbinned time variation for SK-I).  In this combined analysis, SK-II helps expand the 95\% c.l. exclusion from the SK-I-only analysis, mostly along a region dominated by the spectral constraint ($10^{-4}<\tan^2{\theta}<0.4$ and $4\times10^{-5} \textrm{eV}^2<\Delta m^2 <2\times10^{-4} \textrm{eV}^2$).  However, when constraining $^8$B to the SNO NC flux, the SK allowed regions are largely unaffected by the addition of SK-II data.

\subsection{Oscillation Results - SK and Other Solar Experiments}
\begin{figure*}[htbp]
\begin{center}
\includegraphics[scale=0.27]{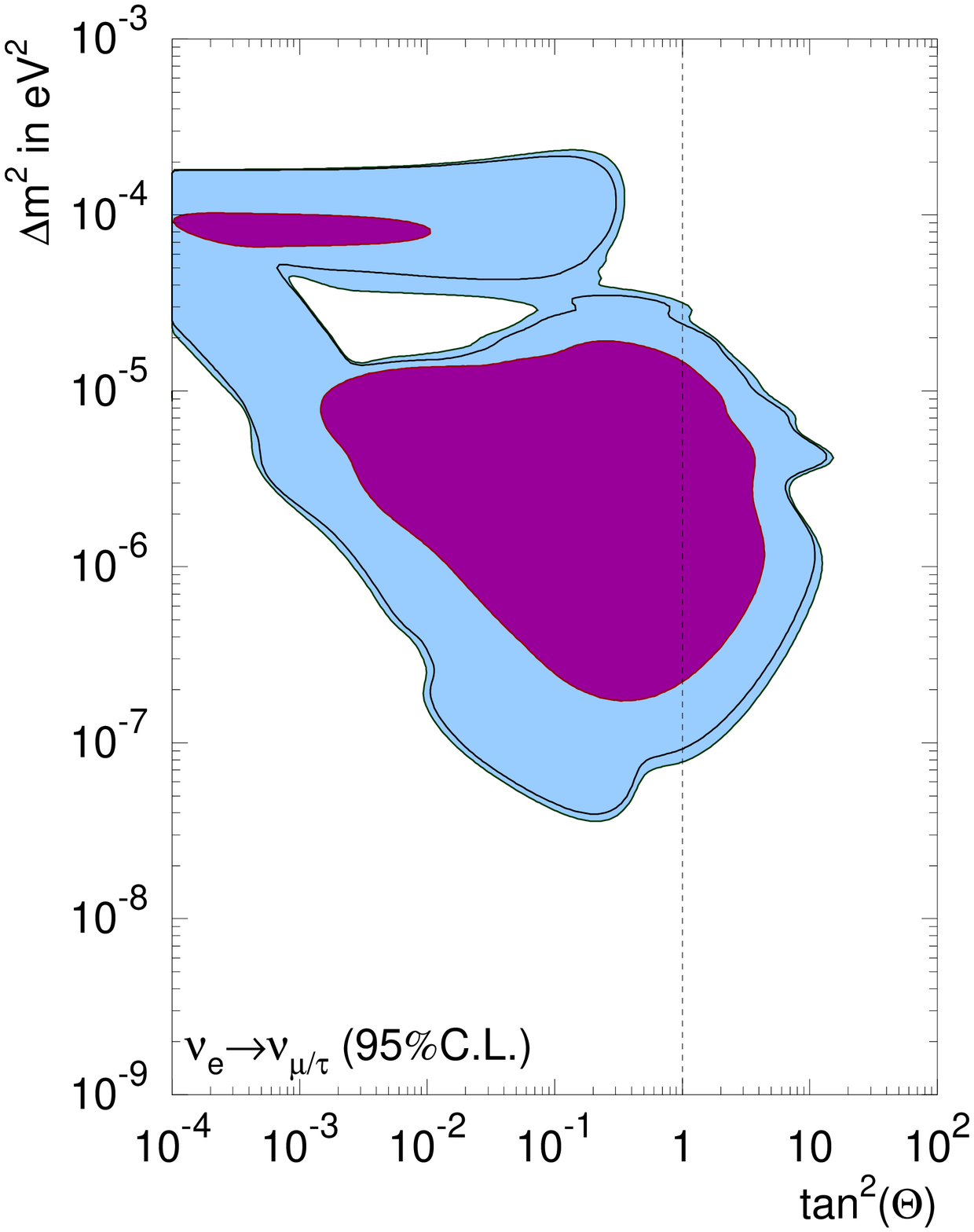}
\includegraphics[scale=0.27]{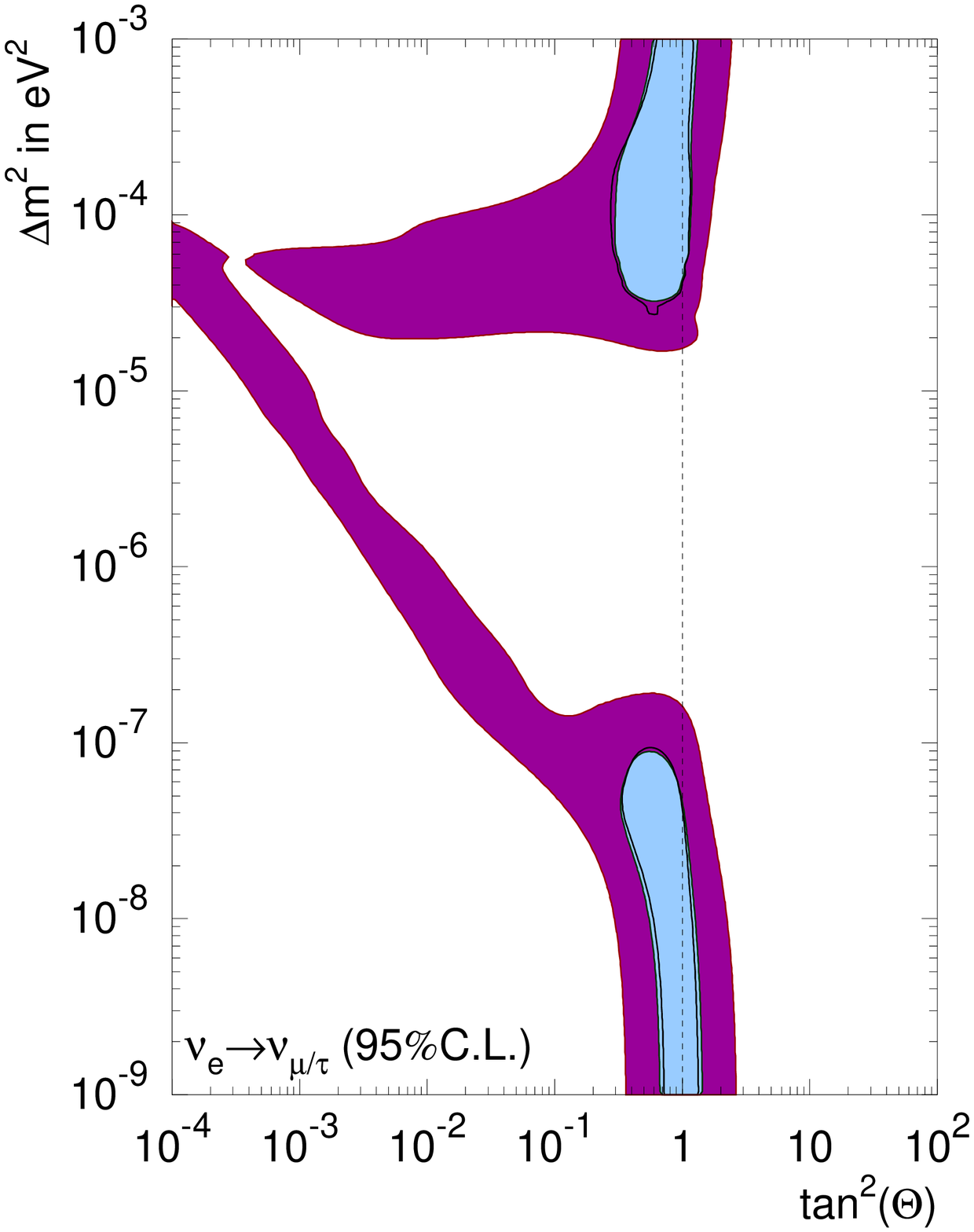}
\includegraphics[scale=0.27]{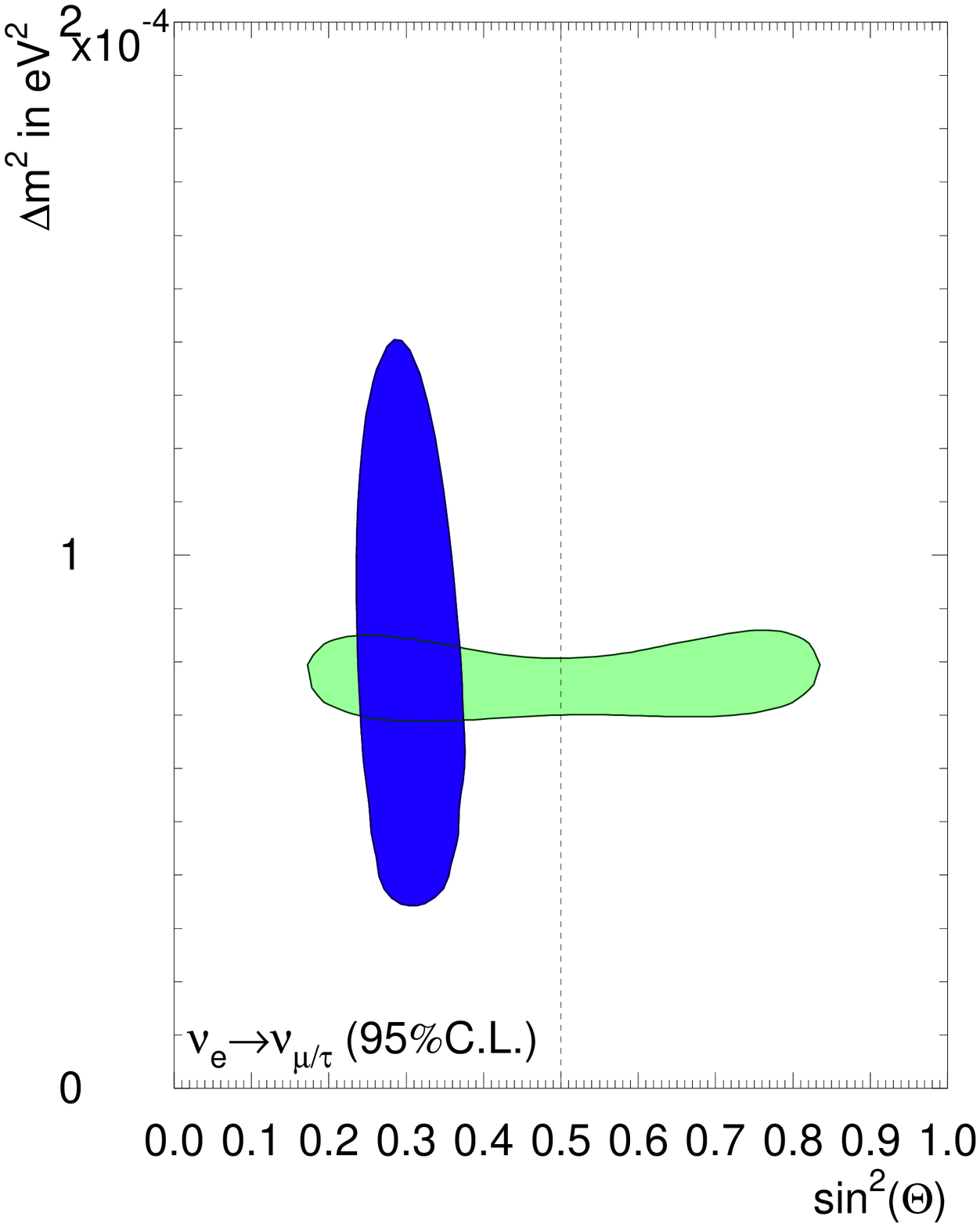}
\caption{The left plot shows SK excluded areas.  The purple region is SK-II and the light blue region is SK-I with SK-II.  The black line shows SK-I only, and evidence of increased exclusion can be seen with the addition of SK-II data.  The center plot shows SK allowed regions (same representative colors as the exclusion contours) with the $^8$B flux constrained to the SNO total flux measurement.  The hep flux is a free parameter.  The right plot shows the SK-I and SK-II combined contour with SNO and radiochemical solar experimental data (blue contour).  The green contour is the KamLAND~\protect\cite{kamland} electron anti-neutrino oscillation result.}
\label{fig:contours}
\end{center}
\end{figure*}

The combination of other solar neutrino experiments such as the SNO and radiochemical results with the SK combined analysis is accomplished by fitting the total CC and NC rates observed by SNO's 306-day pure D$_2$O~\cite{snoncdn} and 391-day salt phases~\cite{snosalt}.  Also, the SNO NC constrained predicted day-night asymmetry for the pure D$_2$O phase is used for added exclusion power.  The radiochemical experiments of Homestake, GALLEX, and SAGE~\cite{othersolar} are then added using the best $^8$B and hep fluxes from the SK-SNO fit.  Figure~\ref{fig:contours} shows the combined solar allowed areas.  The best fit parameter set is $\tan^2\theta = 0.40$ and $\Delta m^2 = 6.03\times 10^{-5} \textrm{eV}^2$ consistent with the SK-I global analysis.

\section{Conclusion}
Super-Kamiokande has measured the solar $^8$B flux to be $(2.38\pm0.05(\textrm{stat.}) ^{+0.16}_{-0.15} (\textrm{sys.}))\times10^6~\textrm{cm}^{-2}\textrm{sec}^{-1}$ during its second phase.  The uncertainties in SK-II are larger than in SK-I but a low analysis threshold of 7 MeV was achieved (7.5 MeV in the day-night variation analysis).  A day-night asymmetry value was observed to be $-0.063 \pm 0.042 (\textrm{stat.}) \pm 0.037 (\textrm{sys.})$ which is consistent with zero and the result from SK-I.  SK-II has brought the total SK time-dependent flux measurement to a length of 9.5 years and this measurement is compared with solar activity in solar cycle 23 resulting in no strong correlation.  In the combined SK-I and SK-II global oscillation analysis, the best fit is found to favor the LMA region at $\tan^2\theta = 0.40$ and $\Delta m^2 = 6.03\times 10^{-5} \textrm{eV}^2$, in excellent agreement with previous solar neutrino oscillation measurements.  SK-I and SK-II agree well, showing no evidence of any systematic effects from the introduction of new methods, blast shields, reduced PMT coverage, etc.

\begin{acknowledgments}
The authors gratefully acknowledge the cooperation of the Kamioka Mining
and Smelting Company. Super-K has been built and operated from
funds provided by the Japanese Ministry of Education, Culture, Sports,
Science and Technology, the U.S. Department of Energy, and the U.S.
National Science Foundation. This work was partially supported by the
Korean Research Foundation (BK21), the Korean Ministry of Science and
Technology, and the National Science Foundation of China.
\end{acknowledgments}

\end{document}